\documentstyle[aps,preprint,graphicx]{revtex}
\title{Vacuum-polarization corrections to the parity-nonconserving $6s-7s$
transition amplitude in $^{133}$Cs.}
\author{
W. R. Johnson\cite{WRJ}, I. Bednyakov and G. Soff}
\address{Instit\"{u}t f\"{u}r Theoretische Physik, Technische
Universit\"{a}t Dresden, Mommenstrasse 13, D-01062 Dresden,
Germany}

\begin{document}
\draft
\tightenlines
%\twocolumn
\maketitle
\begin{abstract}
The dominant one-loop radiative corrections to atomic wave
functions, those associated with vacuum polarization in the
nuclear Coulomb field, are evaluated for the $6s-7s$ parity-nonconserving (PNC)
transition amplitude in $^{133}$Cs. These corrections increase
the size of the PNC amplitude by 0.4\% and, correspondingly, increase
the difference between the experimental value of the weak charge
$Q_W(^{133}{\rm Cs})$ and the value predicted by the
standard model.
\end{abstract}


\section{Introduction}

Parity non-conservation (PNC) in atoms, described in the
standard model of the electroweak interaction by exchange
of $Z$ bosons between bound electrons and nuclear quarks,
leads to nonvanishing electric-dipole matrix elements between
atomic states with the same parity.
The nuclear spin-independent part of PNC matrix elements (arising from the
vector nucleon current) is proportional to a conserved weak charge $Q_W$,
which is sensitive to physics beyond the standard model.

Measurements of the $6s - 7s$ PNC amplitude in $^{133}$Cs,
following the procedure described by Bouchiat and Bouchiat \cite{BB},
were carried out to 2\% accuracy by Gilbert and
Wieman \cite{GW} and, more recently, to 0.3\% accuracy
 by Wood {\it et al.}\ \cite{wood}.
Following the recent measurements, there was a revival of interest
in the associated atomic structure calculations of the PNC
amplitude \cite{DZ,BSJ}. Indeed, Bennett and Wieman \cite{BW}
analyzed differences between experimental and theoretical values
of amplitudes for allowed transitions, polarizabilities, and hyperfine
constants in Cs and concluded that the error in the atomic
structure calculations of the PNC amplitude should be reduced
from 1\%, the value given in Refs.~\cite{DZ,BSJ}, to 0.4\%. Using this
revised estimate of the accuracy of the calculations, they showed
that the experimental value of the weak charge of $^{133}$Cs
differed from the standard-model value \cite{Groom},
\begin{equation}
 Q_W(^{133}{\rm Cs}) = -73.09 \pm 0.03, \label{qw}
\end{equation}
by 2.3~$\sigma$. This difference between
experiment and theory, being one of the two largest
differences reported in the current review of particle physics
\cite{Groom}, suggested the existence of
a second neutral $Z^\prime$ boson \cite{RC,JR,EL}. Implications
of this difference for new physics were also reviewed in 
\cite{RM1} and discussed in \cite{RM2,BK}.  

Breit corrections to the PNC amplitude, which were ignored in
\cite{DZ} and underestimated in \cite{BSJ}, were shown to
decrease the size of the calculated PNC amplitude by a 0.6\% by
Derevianko \cite{der}; this finding was confirmed in
Refs.~\cite{DHJS,KP}. Including Breit corrections reduces the
difference with the standard model to 1.6~$\sigma$ if the 0.4\%
error in the calculations determined in \cite{BW}
is assumed or to 0.9~$\sigma$ if the
more conservative 1\% error given in Refs.~\cite{DZ,BSJ}
is assumed.

\begin{figure}
\centerline{\includegraphics[scale=0.8]{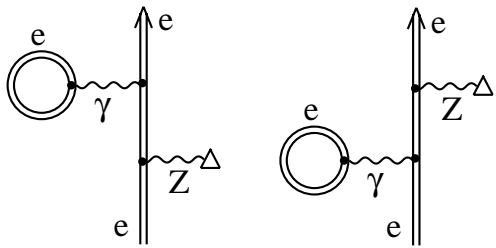}} \caption{One
loop wave function corrections. The double line represents the
electron in the field of the nucleus, the wavy lines represent
$Z$ bosons or photons and the open triangle represents the
$Z$-nucleus vertex. \label{fig1}}
\end{figure}

Radiative corrections to PNC matrix elements
 $\left< w | H_{\rm PNC} | v \right>$ in the strong Coulomb field
of a highly-charged one-electron
ion were considered recently by Bednyakov {\it et al.}
\cite{bed}. These corrections were decomposed into two parts,
radiative corrections to the operator $H_{\rm PNC}$, and
radiative corrections to wave functions $|v\rangle$ and
$|w\rangle$. The former corrections were evaluated for $^{133}$Cs
in Refs.~\cite{LS,MR,MS,S} and are already included in the
theoretical value $Q_W(^{133}{\rm Cs})$ given in Eq.~(\ref{qw}).
The dominant part of the wave function radiative correction,
which arises from the one-loop diagrams Fig.~\ref{fig1}, was
evaluated for $2s-2p_{1/2}$ matrix elements in highly-charged
one-electron ions in Ref.~\cite{bed}. Sushkov, in his
analysis of the Breit corrections \cite{oleg},
suggested that the residual
radiative corrections to PNC amplitudes in many-electron atoms
could be of the same order of magnitude as the Breit corrections.
We examine this suggestion further in the present paper and find
that the one-loop wave-function radiative corrections increase
the size of the PNC amplitude in $^{133}$Cs by 0.4\% and,
correspondingly, increases the difference between the theoretical
and experimental weak charge.  Moreover, we find that the
one-loop wave function correction is insensitive to
electron-electron correlation effects.

\section{Calculation}
One-loop radiative corrections (vacuum polarization corrections)
in an external Coulomb field were considered by Wichmann and Kroll
\cite{WK}, who showed that these corrections simply modify
the electron-nucleus
Coulomb interaction at short range.
The modification of the Coulomb interaction is described,
to leading order in powers of
$Z\alpha$, by the Uehling potential:\cite{UP}
\begin{equation}
\delta V(r) = - \frac{2 \alpha Z}{3 \pi r} \int_1^\infty \!\! dt
\sqrt{t^2-1}
 \left( \frac{1}{t^2} + \frac{1}{2t^4} \right) e^{-2 c t r} . \label{eq1}
\end{equation}
We use atomic units here; $\alpha = 1/137.036\ldots$ is the fine
structure constant and $c \equiv \alpha^{-1}$ is the speed of
light.

In applications to atomic PNC, the Coulomb potential is also modified at
short range by finite nuclear size effects, which are described by
a Fermi-type charge distribution  $\rho(r)$:
\begin{equation}
 \rho(r) = \frac{\rho_0}{1+\exp{[(r-c_{\rm nuc})/a_{\rm nuc}}]} . \label{fs}
\end{equation}
For $^{133}$Cs, we assume that the central radius $c_{\rm nuc}=5.6748$ fm
and that the 10\%--90\% falloff distance is $t_{\rm nuc}=2.3$ fm,
corresponding to $a_{\rm nuc}=0.523$ fm and to a root-mean-square radius of
the nuclear charge distribution $R_{\rm rms}= 4.807$ fm.
As shown by Fullerton and Rinker \cite{FR}, the Uehling
potential can be generalized as follows to accommodate the charge
distribution $\rho(r)$ of an extended nucleus:
\begin{eqnarray}
\lefteqn{\delta V(r)
 = - \frac{2\alpha^2}{3r} \int_0^\infty \! dx\, x\, \rho(x)
\int_1^\infty dt
\sqrt{t^2-1}\ \times}
\hspace{4em}\nonumber\\
&& \left( \frac{1}{t^3} + \frac{1}{2t^5} \right)
\left( e^{-2 c t |r-x|}-e^{-2 c t (r+x)} \right) . \label{vp}
\end{eqnarray}

Corrections to Dirac-Coulomb energies
for $1s$, $2s$ and $2p$ electrons in one-electron ions
obtained by solving the Dirac equation in the composite
nuclear + Uehling potential discussed above
are in found to be in close agreement with sums of the Uehling,
Uehling--finite nuclear size, and finite nuclear size
corrections obtained perturbatively in \cite{JS}.
A comparison of the present
results with those from \cite{JS} are given in Table~\ref{tab1}.
Uehling corrections to $ns$ and $np_{1/2}$ levels of Cs
(the levels of primary interest in PNC calculations in Cs)
are about twice as large as finite nuclear size corrections and
have opposite signs as shown in the table for levels with
$n$=1 and 2.

We carry out two calculations of the $6s-7s$ PNC amplitude
in the modified potential, both leading to precisely the same relative
correction to the PNC amplitude. The first of these calculations,
is done at the ``weak'' Dirac-Hartree-Fock (DHF) level of approximation.
The perturbation
$\delta\phi_v^{\rm HF}$
to a valence electron wave function $\phi_v^{\rm HF}$ induced by the {\em weak} interaction
$h_{\rm PNC}$ satisfies the inhomogeneous DHF equation
\begin{equation}
 \left( h_0 + V^{\rm HF} -\epsilon_v^{\rm HF} \right)
 \delta{\phi}_v^{\rm HF} = -h_{\rm PNC} \phi_v^{\rm HF} .
\label{EHF}
\end{equation}
In this equation, $V_{\rm HF}$ is the HF potential of the closed xenon-like core
and $\epsilon_v^{\rm HF}$ is the eigenvalue of the unperturbed DHF equation.
The perturbed DHF equations are solved to give the
$\delta{\phi}_{6s}^{\rm HF}$
and $\delta{\phi}_{7s}^{\rm HF}$.  The PNC amplitude is
then given by the sum of two terms:
\begin{equation}
E_{\rm PNC} =   \langle \phi_{7s}^{\rm HF} | D | \delta{\phi}_{6s}^{\rm HF} \rangle
+ \langle \delta{\phi}_{7s}^{\rm HF} | D | \phi_{6s}^{\rm HF} \rangle\, , \label{sum}
\end{equation}
where $D$ is the dipole operator. This calculation leads to a value of the
PNC amplitude that is 20\% lower than the final correlated value found in
Refs.~\cite{DZ,BSJ,KP}. In the top panel of Table~\ref{tab2}, we show each of the two terms making
up the sum in Eq.~(\ref{sum}) determined with and without the modified Uehling
potential $\delta V$ from Eq.~(\ref{vp}). The one-loop correction is seen to increase the
size of each terms and their sum by 0.41\%.

The second calculation is done at the ``weak'' random-phase approximation (RPA)
level of approximation, in which the
class of correlation corrections associated with weak perturbations of the
core orbitals  are included in the calculation. This calculation leads to a value
of the PNC amplitude that is 3\% larger than the final correlated value
of the amplitude given in \cite{DZ,BSJ,KP}.
This class of correlation corrections is included by solving
\begin{equation}
\left( h_0 + V^{\rm HF} -\epsilon_v^{\rm HF} \right)
\delta \phi^{\rm RPA}_v =  - \left[ h_{\rm PNC}
+  V^{\rm HF}_{\rm PNC} \right]  \phi_v^{\rm HF} ,
\label{EBr}
\end{equation}
where $V^{\rm HF}_{\rm PNC}$ is the weak correction to the HF potential.
The resulting PNC amplitude is given by
\begin{equation}
E_{\rm PNC} =
 \langle \phi_{7s}^{\rm HF} | D | \delta \phi^{\rm RPA}_{6s} \rangle
+ \langle \delta \phi^{\rm RPA}_{7s} | D | \phi_{6s}^{\rm HF} \rangle\, . \label{sumb}
\end{equation}
In the lower  panel of Table~\ref{tab2}, we give values of the two terms making
up the sum in Eq.~(\ref{sumb}) with and without the  Uehling
corrections.  Again, the one-loop correction is seen to increase the
size of each of the two terms and their sum by 0.41\%.

From the two calculations above, it is apparent that the short
range vacuum-polarization corrections are independent
of electron-electron correlation.  The situation is similar to
that found for the short range nuclear ``skin'' correction arising from
the difference between the neutron and proton radius of the nucleus
\cite{wel,james},
which also leads to a correlation-independent correction to the PNC amplitude.

\section{Conclusion}
One-loop radiative corrections to the electron
wave functions from vacuum-polarization in the nuclear field
are evaluated.
The resulting wave functions are used to calculate the PNC amplitude
for the $6s-7s$ PNC transition
in $^{133}$Cs, leading to a 0.4\% increase the size of the amplitude.
This correction is found to be correlation independent.

An average of the three most accurate calculations\cite{DZ,BSJ,KP},
of the $6s-7s$ amplitude,
taking account of Breit corrections and one-loop radiative corrections
gives
\[
E_{\rm PNC} = -9.057\ \pm 0.037 \ iea_0\times 10^{-12} (-Q_W/N),
\]
were we use the estimate from
\cite{BW} for the error in the calculations. Combining this
calculated amplitude
with the experimental value of the amplitude from \cite{wood}
leads to an experimental weak charge
\[
Q^{\rm expt}_W(^{133}{\rm Cs}) = -72.12 \pm (0.28)_{\rm expt} \pm (0.34)_{\rm theor}
\]
This value differs by 2.2~$\sigma$ from the standard model.
If we assume a 1\% error in the theoretical amplitude, then
the theoretical component of the error in $Q^{\rm expt}_W$ is increased to
$\pm (0.74)_{\rm theor}$
and the difference with the standard model becomes 1.2~$\sigma$.

We have considered only the dominant one-loop wave-function
radiative corrections in the above calculation and ignored the
higher-order $\alpha Z$ corrections to the Uehling potential
discussed by Wichmann and Kroll \cite{WK}.  The  Wichmann-Kroll
corrections can be easily estimated and are expected to have a
negligible effect on the present result.  Moreover, the
``vertex'' radiative correction to the PNC amplitude, which
contributes about 0.1\% to the Standard Model value of $Q_W$, is
evaluated in Refs.~\cite{LS,MS} using free-particle propagators.
This radiative correction should be redone using Coulomb-field
propagators.  If Coulomb-field effects were to change the vertex
correction by 100\% (probably a gross overestimate) the
Standard Model value of $Q_W$ would change by only 0.1\%.
We therefore expect that further changes to the value of $Q_W$
from the strong-field corrections to the operator $H_{\rm PNC}$
discussed in \cite{bed} will be smaller than 0.1\%.

 \acknowledgements The work
of WRJ was supported in part by an Alexander von Humboldt award
from the Federal Republic of Germany and in part by the U. S.
National Science Foundation under grant No.\ 99-70666. WRJ owes a
debt of gratitude to O. Sushkov for helpful discussions. GS and
IB acknowledge fruitful discussions with L. Labzowsky and support by
the BMBF and GSI (Darmstadt).
\newpage

\begin{table}
\caption{Differences between the present one-electron Dirac energies
for Cs ($Z$=55) in the finite nucleus
plus Uehling potential and Dirac-Coulomb energies  ($\delta E$)
are compared with
perturbative values of Uehling (Uehl), Uehling--finite nuclear size
(Uehl-FS), and finite nuclear size (FS) corrections to one-electron
energy levels given in \protect\cite{JS}. Units:
$\alpha m c^2 (\alpha Z)^4 /(\pi n^3)$ \label{tab1}}
\begin{tabular}{clllll}
\multicolumn{1}{c}{} &
\multicolumn{1}{c}{Present} &
\multicolumn{4}{c}{Ref.~\protect\cite{JS}} \\
\cline{3-6}
\multicolumn{1}{c}{State} &
\multicolumn{1}{c}{$\delta E$} &
\multicolumn{1}{c}{Uehl} &
\multicolumn{1}{c}{Uehl-FS} &
\multicolumn{1}{c}{FS} &
\multicolumn{1}{c}{Total}\\
\hline
 $1s_{1/2}$ & -0.1419  & -0.2584  & 0.0010(1)  &  0.1159   &  -0.1415(1)\\
 $2s_{1/2}$ & -0.1554  & -0.2901  & 0.0011(1)  &  0.1339(1)&  -0.1551(2)\\
 $2p_{1/2}$ & -0.0100  & -0.0145  & 0.0000(1)  &  0.0045   &  -0.0100(1)\\
 $2p_{3/2}$ & -0.0016  & -0.0016  & 0.0000     &  0.0000   &  -0.0016
\end{tabular}
\end{table}

\begin{table}
\caption{One-loop corrections to the weak HF-PNC amplitude for the $6s-7s$
PNC amplitude for $^{133}$Cs are shown in the upper panel and
to the weak RPA-PNC amplitude in the lower panels. Values are listed without
and with corrections from Eq.~(\ref{vp}).
 Units: $iea_0\times 10^{-12} (-Q_W/N)$ \label{tab2}}
\begin{tabular}{cccc}
\multicolumn{1}{c}{Type} &
\multicolumn{1}{c}{$\langle \phi_{7s}|D|\delta \phi_{6s}\rangle$} &
\multicolumn{1}{c}{$\langle \delta \phi_{7s}|D|\phi_{6s}\rangle$} &
\multicolumn{1}{c}{$E_{\rm PNC}$}\\
\hline
\multicolumn{4}{l}{Weak HF approximation:\rule[0pt]{0em}{1.2em} } \\
DHF              &  2.749183&   -1.014387&   -7.394685\\
DHF+ $\delta V$  &  2.760414&   -1.018581&   -7.425391\\[0.1pc]
$\Delta$ (\%)    &      0.41&        0.41&        0.41\\[0.3pc]
\multicolumn{4}{l}{Weak RPA approximation:} \\
RPA              &  3.457036&   -1.272562&   -9.268581\\
RPA+ $\delta V$  &  3.471169&   -1.277834&   -9.307166\\[0.1pc]
$\Delta$ (\%)    &      0.41&        0.41&        0.41
\end{tabular}
\end{table}

\end{document}